\documentstyle[aps]{revtex}
\begin{document}
\title{Fundamental equations for the gravitational and electromagnetic
perturbations of a charged black hole}
\author{Zolt{\'{a}}n Perj{\'{e}}s}
\address{KFKI Research Institute for Particle and Nuclear Physics,\\
H--1525, Budapest 114, P.O.B.\ 49, Hungary\\
{\tt e-mail: perjes@rmki.kfki.hu}ÿ}
\date{\today }
\maketitle

\begin{abstract}
A pair of wave equations is presented for the gravitational and
electromagnetic perturbations of a charged black hole. One of the equations
is uncoupled and determines the propagation of the electromagnetic
perturbation. The other is for the propagation of the shear of a principal
null direction and has a source term given by the solution of the first
equation. This result is expected to have important applications in
astrophysical models.

\end{abstract}

The description of gravitational, electromagnetic and Weyl neutrino field
perturbations in terms of a separable master equation for an uncharged black
hole\cite{Teukolsky} has found a wide range of applications, which include
the stability of black holes, the characteristics of gravitational and
electromagnetic radiation and observations of distant relativistic objects.
However, despite numerous attempts\cite
{Fackerell,Chandrasekhar,Chitre,Crossman} to find a corresponding
description for of the fields for charged Kerr-Newman black holes, success
eluded this strive until now.

Meanwhile, the discoveries of new classes of bright astrophysical X-ray
sources by, among others, the {\it XMM} and {\it Chandra} missions\cite{XMM}
have intensified the search for models involving charged black holes\cite
{Znajek}.

The purpose of this Letter is to present a pair of separable wave equations
for{\it \ the electromagnetic and the gravitational perturbation components}
of the charged black hole. These wave equations arise in a linear
approximation to the coupled Einstein-Maxwell equations. It is hoped that
the perturbation formalism presented in this paper will find applications in
the stability theory of charged black holes, in the theory of relativistic
sources of electromagnetic and gravitational radiation, and in models of
astrophysical X-ray sources.

In a recent paper\cite{Perjes}, the stationary perturbations of the charged
Kerr black hole have been shown to obey a pair of simple wave equations.
This structure arises in a gauge adapted to the complex bivector $\nabla
_aK_b+i\left( \nabla _aK_b\right) ^{*}$ where $K=\partial /\partial t$ is
the timelike Killing vector of the black-hole space-time and an asterisk
denotes the dual bivector. The procedure for the stationary perturbations
provides a convenient guidance for the generic description. (The full
details will be given in a subsequent paper\cite{bhpart}).

The Kerr-Newman space-time with mass ${\rm m}$, rotation parameter $a$ and
electric charge ${\rm e}$ has the metric\cite{Kerr,Newman} 
\begin{equation}
ds^2=\left( 1-%
{\textstyle {2{\rm m}r-{\rm e}^2 \over \zeta \overline{\zeta }}}
\right) \left( dt-a\sin ^2\vartheta d\varphi \right) ^2+2\left( dt-a\sin
^2\vartheta d\varphi \right) \left( dr+a\sin ^2\vartheta d\varphi \right)
-\zeta \overline{\zeta }\left( d\vartheta ^2+\sin ^2\vartheta d\varphi
^2\right)  \label{ds2}
\end{equation}
and four-potential 
\begin{equation}
A=-%
{\textstyle {{\rm e}r \over \zeta \overline{\zeta }}}
\left( dt-a\sin ^2\vartheta d\varphi \right)  \label{A}
\end{equation}
where 
\begin{equation}
\zeta =r-ia\cos \vartheta .
\end{equation}

Previous approaches to the subject use the Kinnersley tetrad\cite{Kinnersley}%
, the two real null vectors of which are chosen to lie in the double
principal null directions of the curvature. The first important hint from
the stationary study\cite{Perjes} is to use, instead, a gauge related to the
Killing vector. Indeed, we employ a Newman-Penrose approach\cite{NP} where
we choose the tetrad such that the null vector $\ell $ points in a principal
null direction, and the null vector $n$ is in the intersection of the
2-plane of $\ell $ and the Killing vector $K$ with the null cone. This
amounts to selecting the Newman {\it et al.} tetrad. In the present
coordinate system, this is 
\begin{eqnarray}
D &\equiv &\ell ^a%
{\textstyle {\partial \over \partial x^a}}
=%
{\textstyle {\partial \over \partial r}}
\nonumber \\
\Delta &\equiv &n^a%
{\textstyle {\partial \over \partial x^a}}
=%
{\textstyle {1 \over 2}}
\left[ 
{\textstyle {\left( \zeta +\overline{\zeta }\right) {\rm m}-{\rm e}^2 \over \zeta \overline{\zeta }}}
-1\right] 
{\textstyle {\partial \over \partial r}}
+%
{\textstyle {\partial \over \partial t}}
\\
\delta &\equiv &m^a%
{\textstyle {\partial \over \partial x^a}}
=%
{\textstyle {1 \over 2^{1/2}\overline{\zeta }}}
\left[ 
{\textstyle {\partial \over \partial \vartheta }}
+%
{\textstyle {i \over \sin \vartheta }}
{\textstyle {\partial \over \partial \varphi }}
-ia\sin \vartheta \left( 
{\textstyle {\partial \over \partial r}}
-%
{\textstyle {\partial \over \partial t}}
\right) \right]  \nonumber \\
\overline{\delta } &\equiv &\overline{m}^a%
{\textstyle {\partial \over \partial x^a}}
\nonumber
\end{eqnarray}
and the Maxwell tensor components are 
\begin{eqnarray}
\Phi _0 &\equiv &F_{ab}\ell ^am^b=0  \nonumber  \label{Phis} \\
\Phi _1 &\equiv &%
{\textstyle {1 \over 2}}
F_{ab}\left( \ell ^an^b+\overline{m}^am^b\right) =\frac{{\rm e}}{%
2^{1/2}\zeta ^2}  \label{Phis} \\
\Phi _2 &\equiv &F_{ab}\overline{m}^an^b=\frac{i{\rm e}a\sin \vartheta }{%
\zeta ^3}.  \nonumber
\end{eqnarray}
With this choice of the tetrad for the charged Kerr black hole, two
components of the Weyl curvature $C_{abcd}$ vanish, $\Psi _0\equiv
-C_{abcd}\ell ^am^b\ell ^cm^d=0$ and $\Psi _1\equiv -C_{abcd}\ell ^an^b\ell
^cm^d=0$.

In the generic case, the Weyl tensor of the {\em perturbed} space-time has
four distinct principal null directions which are pairwise bundled around
the double null directions of the Kerr-Newman space-time. We choose the
perturbed tetrad once again such that the spinor $o^A$ is a principal spinor
of the Weyl curvature: 
\begin{equation}
\Psi _0=0.  \label{pnd}
\end{equation}
This choice, though convenient for the present purposes, is not the only one
that can be used. With other choices of the perturbed tetrad, the function $%
\Psi _0$ would carry information about our failure to align $\ell $ with a
principal null direction. In Ref.\cite{Perjes}, both the `{\it eigenray
condition}' and a gauge adapted to the perturbed Maxwell field have been
described. The generalization of the former gauge condition to an arbitrary
perturbed space-time differs from (\ref{pnd}) by perturbation terms
involving the shear $\sigma $.

When such a choice of the tetrad is made to full precision, the direction of
the vector $\ell $ is fixed up to a discrete group of transformations
permuting the bundled principal null directions. In the present case we are
interested in a description keeping only first-order corrections. Under the
infinitesimal dyad transformation 
\begin{equation}
o^A\rightarrow o^A+b\iota ^A,\qquad \iota ^A\rightarrow \iota ^A,
\label{inftr}
\end{equation}
where $b$ is an arbitrary but small complex multiplier function such that
higher powers of $b$ are negligible, the quantity $\Psi _0$ transforms as
follows: 
\begin{equation}
\Psi _0\rightarrow \Psi _0+4b\Psi _1.
\end{equation}
Both $b$ and the curvature quantity $\Psi _1$ are small. Thus we find
serendipitiously that, to the required accuracy, the spinor $o^A$ remains a
principal spinor of the curvature under the transformations (\ref{inftr})!
We can use this gauge freedom, in a similar fashion as Crossman\cite
{Crossman} does, to eliminate all coupling terms from the wave equation 
\begin{equation}
\Box _1\phi =0  \label{Phieq}
\end{equation}
where the wave operator is 
\begin{eqnarray}
\Box _s &=&{\bf \Delta }^{-s}%
{\textstyle {\partial \over \partial r}}
{\bf \Delta }^{s+1}%
{\textstyle {\partial \over \partial r}}
+%
{\textstyle {1 \over \sin \vartheta }}
{\textstyle {\partial \over \partial \vartheta }}
\sin \vartheta 
{\textstyle {\partial \over \partial \vartheta }}
+s\left( 1-s%
{\textstyle {\cos ^2\vartheta  \over \sin ^2\vartheta }}
\right)  \nonumber \\
&&{\cal +}\left[ 2a\left( 
{\textstyle {\partial \over \partial t}}
-%
{\textstyle {\partial \over \partial r}}
\right) +%
{\textstyle {1 \over \sin ^2\vartheta }}
\left( 
{\textstyle {\partial \over \partial \varphi }}
+2is\cos \vartheta \right) \right] 
{\textstyle {\partial \over \partial \varphi }}
\label{waveop} \\
&&+a^2\sin ^2\vartheta 
{\textstyle {\partial ^2 \over \partial t^2}}
-2(r^2+a^2)%
{\textstyle {\partial ^2 \over \partial r\partial t}}
-2\left[ (s+2)r+ia\cos \vartheta \right] 
{\textstyle {\partial \over \partial t}}
,  \nonumber
\end{eqnarray}
the electromagnetic perturbation function $\phi $ is defined by 
\begin{equation}
\Phi _0={\rm e}\phi  \label{phidef}
\end{equation}
and ${\bf \Delta }=r^2-2{\rm m}r+a^2+{\rm e}^2$.

The second wave equation can be obtained by essentially repeating the steps
made in Ref.\cite{Perjes} for deriving the master equation of the
gravitational perturbation components. In the resulting equation, each term
on the left contains the small function 
\begin{equation}
\psi =\frac \sigma {\zeta ^2}  \label{psidef}
\end{equation}
and each term on the right contains either a $\phi $ or the first-order spin
coefficient $\kappa \equiv \ell _{a;b}m^a\ell ^b$. When inserting the
unperturbed values of the operators and factors, neither the $\psi $ terms,
nor the $\phi $ terms cancel but those with $\kappa $ do. We get the wave
equation 
\begin{equation}
\Box _2\psi =\frac 1{\overline{\zeta }^2}J\phi .  \label{psieq}
\end{equation}
Thus the solution $\phi $ of Eq. (\ref{Phieq}) will provide the source
function $J\phi $ for the equation for $\psi .$ The source term is a
functional of the field $\phi $ containing up to second derivatives, with
the operator 
\begin{eqnarray}
J &=&\left( {\rm e}^2-{\rm m}\zeta \right) \left( 
{\textstyle {\partial \over \partial \vartheta }}
+ia\sin \vartheta 
{\textstyle {\partial \over \partial t}}
+%
{\textstyle {i \over \sin \vartheta }}
{\textstyle {\partial \over \partial \varphi }}
-%
{\textstyle {\cos \vartheta  \over \sin \vartheta }}
\right) 
{\textstyle {\partial \over \partial r}}
\\
&&-%
{\textstyle {1 \over \overline{\zeta }}}
\left[ {\rm e}^2+{\rm m}\left( 2\overline{\zeta }-\zeta \right) \right]
\left( 
{\textstyle {\partial \over \partial \vartheta }}
+ia\sin \vartheta 
{\textstyle {\partial \over \partial t}}
+%
{\textstyle {i \over \sin \vartheta }}
{\textstyle {\partial \over \partial \varphi }}
-%
{\textstyle {\cos \vartheta  \over \sin \vartheta }}
\right)  \nonumber \\
&&+%
{\textstyle {1 \over \overline{\zeta }}}
\left[ {\rm e}^2-{\rm m}\left( 2\overline{\zeta }+\zeta \right) \right]
ia\sin \vartheta 
{\textstyle {\partial \over \partial r}}
.  \nonumber
\end{eqnarray}

When an additional matter source is present, both equations (\ref{Phieq})
and (\ref{psieq}) acquire a source term proportional to the corresponding
stress-energy tensor $T_{ab}.$ One can treat these equations by expanding
the homogeneous solutions in quasi-normal modes with energy $\omega $ and
helicity $m$\cite{Teukolsky}: 
\begin{equation}
\Phi _0=\int d\omega \sum_{l,m}R\left( r\right) S_l^m\left( \vartheta
\right) e^{i(m\varphi -\omega t)}.
\end{equation}
Here the radial function $R\left( r\right) $ and the angular function $%
S_l^m\left( \vartheta \right) $ satisfy the ordinary differential equations,
respectively, 
\begin{equation}
\left\{ {\bf \Delta }^{-s}%
{\textstyle {\partial \over \partial r}}
{\bf \Delta }^{s+1}%
{\textstyle {\partial \over \partial r}}
+2i\left[ (r^2+a^2)\omega -am\right] 
{\textstyle {\partial \over \partial r}}
+2\omega \left[ i(s+2)r+am\right] -{\cal A}-\omega ^2a^2\right\} R=0
\label{radeq}
\end{equation}
\begin{equation}
\left[ 
{\textstyle {1 \over \sin \vartheta }}
{\textstyle {\partial \over \partial \vartheta }}
\sin \vartheta 
{\textstyle {\partial \over \partial \vartheta }}
+\left( \omega a\cos \vartheta -1\right) ^2{\cal -}\left( 
{\textstyle {m+s\cos \vartheta  \over \sin \vartheta }}
\right) ^2+s-1+{\cal A}\right] S_l^m=0  \label{angeq}
\end{equation}
and ${\cal A}=\Lambda -\omega ^2a^2$ where $\Lambda $ is the separation
constant of the kernel of operator $\Box _s$.

An advantageous feature of the coordinate system used here is that the
metric (\ref{ds2}) remains regular on the null hypersurfaces ${\bf \Delta }%
=0.$ Although the radial equation (\ref{radeq}) has a singularity on the
horizon (the outer solution of ${\bf \Delta }=0$), one can choose the
boundary conditions for the wave equations (\ref{radeq}) and (\ref{angeq})
on the horizon such that the perturbations $\phi $ and $\psi $ are regular.

Methods for solving propagation equations, closely related to (\ref{radeq})
and (\ref{angeq}), with source terms have been presented in \cite{Teukolsky}%
. Separability of the master equations is not, however, the only property by
which solution can be achieved. Chrzanowski and Misner\cite{Chrzanowski}
write down the solution for the uncharged black hole using the method of
asymptotically factorized Green's functions.

Given a solution of Eqs. (\ref{Phieq}) and (\ref{psieq}), the perturbation
functions $\sigma $ and $\Phi _0$ are available from the respective simple
relations (\ref{psidef}) and (\ref{phidef}). One can next compute the spin
coefficient $\kappa $ by a method quite similar to that described in \cite
{Perjes}. Continuing the step-by-step integration procedure, we have an
algorithm for systematically getting the full description of the perturbed
space-time.

\acknowledgments{\ 
}

I thank M\'{a}ty\'{a}s Vas\'{u}th for discussions. This work has been
supported by the OTKA grant T031724.

\end{document}